\documentstyle[aps,ifthen,graphicx,subfigure]{revtex}
\title{Finite temperature phase transition in the two-dimension 
Randomly Coupled Ferromagnet }
\author{N. Lemke\footnote{lemke@if.ufrgs.br lemke\@exatas.unisinos.tche.br}}
\address{Centro de Ci\^encias Exatas e da Terra -- Unisinos \\
Av. Unisinos, 950 \\
93022-000 -- S\~ao Leopoldo -- RS }
\author{I. A. Campbell\footnote{campbell@lps.u-psud.fr}}
\address{
Laboratoire de Physique des Solides  \\
Universit\'e Paris Sud, 91405 Orsay, France }

\newcommand{\version}{condmat}

\begin{document}

\maketitle

\begin{abstract}
We show using extensive simulation results and physical arguments that an
Ising system on a two dimensional square lattice, having interactions of
random sign between first neighbors and ferromagnetic interactions between
second neighbors, presents a phase transition at a non-zero temperature.

\end{abstract}

\section*{Introduction and definition of the model}

The Edwards-Anderson Model for spin-glasses is one of the most
intensively studied models in the domain of disordered systems. It
combines a very simple formulation and rich behavior. However in its
standard form it does not possess a finite temperature transition in
dimension two \cite{bhatt1988}, so the study of finite temperature
spin glass phase transitions is restricted to three dimensions or
above. This is unfortunate as dimension two has many advantages,
including the facility for direct visualization.

The Random Field Ising Model (RFIM) proposed by Imry and Ma
\cite{imry1975} is also a very important model for disordered systems.
Unfortunately this model does not present a phase transition in  two
dimensions \cite{imbrie1984}, implying that once more we have to go to
higher dimensions to study a phase transition.

These results seem to imply that no intrinsically two dimensional disordered system
present a finite transition temperature. This statement has been
questioned in recent years and some disordered systems have been
proposed for a finite temperature phase transition in two dimensions
\cite{shirakura1997,pasquini1997}.

We introduced \cite{lemke1996b} a model for Ising spins on a square
lattice where second neighbors are coupled ferromagnetically with an
interaction strength $J$, and where there is a random near neighbor
coupling of strength $\pm \lambda J$. The model is described by the
following Hamiltonian:
\begin{equation}
\label{def.eq}
{\cal H} =\sum_{\langle i,j \rangle} - S_iJ S_j + \sum_{\{ i,j\} }
S_iJ_{ij}S_j \end{equation}
where $\langle i,j \rangle $ means a sum over second neighbors, $\{ i,j\}$
over first neighbors and $J_{ij}=\pm \lambda J$. We have decided to call this model the  
Randomly Coupled Ferromagnet (RCF).  We presented simulations
which indicated the presence of a phase transition at a finite temperature
$T_c$ near $2$ (in units of $J$) for $\lambda < 1$. Similar behavior were
also found on the analogous XY model \cite{jain1998}.

The conclusions we drew for the Ising version of this model were
contested in a paper by Parisi, Ruiz-Lorenzo, and Stariolo
\cite{parisi1998}. These authors carried out simulations to larger
sizes, up to $L=48$. They interpreted their data in terms of size
dependent crossovers at low temperatures, successively between
staggered ferromagnet, spin-glass like, and paramagnetic phases as
size $L$ increases. In \cite{parisi1998} a picture for the low
temperature phase was proposed implying that at large sizes the system
is equivalent to a standard two dimension spin-glass, having no true
finite temperature phase transition.

Here we give arguments leading to a different picture for the low
temperature phase. Simulations are presented which clearly indicate a
phase transition at a finite temperature for a wide range of
$\lambda$.

\section*{General Discussion of the model}

When $\lambda =0$ we simply have two independent sub-lattices $A$ and $B$
that order ferromagnetically and independently at the Onsager value of the
Curie temperature, $T_c=2.26\ldots$. Below this temperature each sub-lattice
has its overall magnetization either up or down; thus there are four
different degenerate ground states.

For finite values of $\lambda$, each sub-lattice will exert a random field
on the other, so we expect that for a large enough lattice the long range
ferromagnetic order in each sub-lattice will be destroyed, following the
Imry-Ma argument for the two dimensional RFIM
\cite{imry1975}. On each sub-lattice the system will be broken up into
several different locally ferromagnetic domains whose size will be roughly
given by
\begin{equation}
R_c\sim \exp\left[\frac{C}{\lambda^2} \right], 
\end{equation} where $C$ is
a constant \cite{parisi1998,binder1983,seppala1998}.

However here in contrast to the RFIM the ``effective random fields'' are not
fixed once and for all but fluctuate in time as the spins in the other
sub-lattice relax. The crucial point is: are the domains ``stable'' in
time below some critical temperature? Alternatively, are they in perpetual
motion at all finite temperatures, so that after a sufficient time all
memory of an initial equilibrium spin configuration will be wiped out, even
in the thermodynamic limit? In reference \cite{parisi1998} it is proposed
that each of these domains can be regarded as a ``super-spin'' having finite
random interactions with its neighbors. So the system will behave as
ferromagnet for small scales and as a spin glass for larger scales. This
picture implies that in the thermodynamical limit the system will only
present a phase transition at $T=0$.

We argue that this picture is incorrect, since in fact the ``super-spins'' do
not behave as na\"{\i}vely expected. Let us first go to the RFIM limit. In
figure \ref{fixed.fig} we present the results of a simulation at $\lambda =
0.5$ where the spins on sub-lattice $A$ are all frozen up, and those on
sub-lattice $B$ can evolve following the Hamiltonian (\ref{def.eq}).
Sub-lattice $A$ is reduced basically to one super-spin, and the figure also
represents a snapshot of the $B$ sub-lattice configuration after a long
anneal at temperature $T=1.5$, together with the time dependence of the
auto correlation function after anneal, $q_B(t)$, at the same temperature
for the $B$ spins. $q_B(t)$ is defined as:

\begin{equation}
\label{q.eq}
q_B(t) =\lim_{t_w\to \infty}\sum_{i\in B} [ S_i(t_w)S_i(t_w+t) ]
\end{equation}
where $[ \ldots ]$ represent a configurational average. As we can see
$q_B(t)$ quite rapidly reaches an asymptotic value, which is about 0.85 at
this temperature. The system can be exactly described by the Hamiltonian:
\begin{eqnarray}
\label{rf.eq}
{\cal H}(S_i=1 |\, i\in A) &=& \sum_{<i,j>} J S_i S_j +
\sum_{i,j}S_iJ_{ij}S_j \\ &=& \sum_{<i,j>} J S_i S_j + \sum_{i}h_i S_i
\end{eqnarray}
where $h_i$ is a random field. This is precisely the RFIM hamiltonian. The
Imbrie result \cite{imbrie1984} implies that for a infinite system $q_B(t)$
will tend to definite positive value for all temperatures. 

\ifthenelse{\equal{\version}{condmat}}{  
\begin{figure}
\begin{center}
\subfigure[]{\includegraphics[bb= 0 0 600 800, scale=0.3, angle= 270]{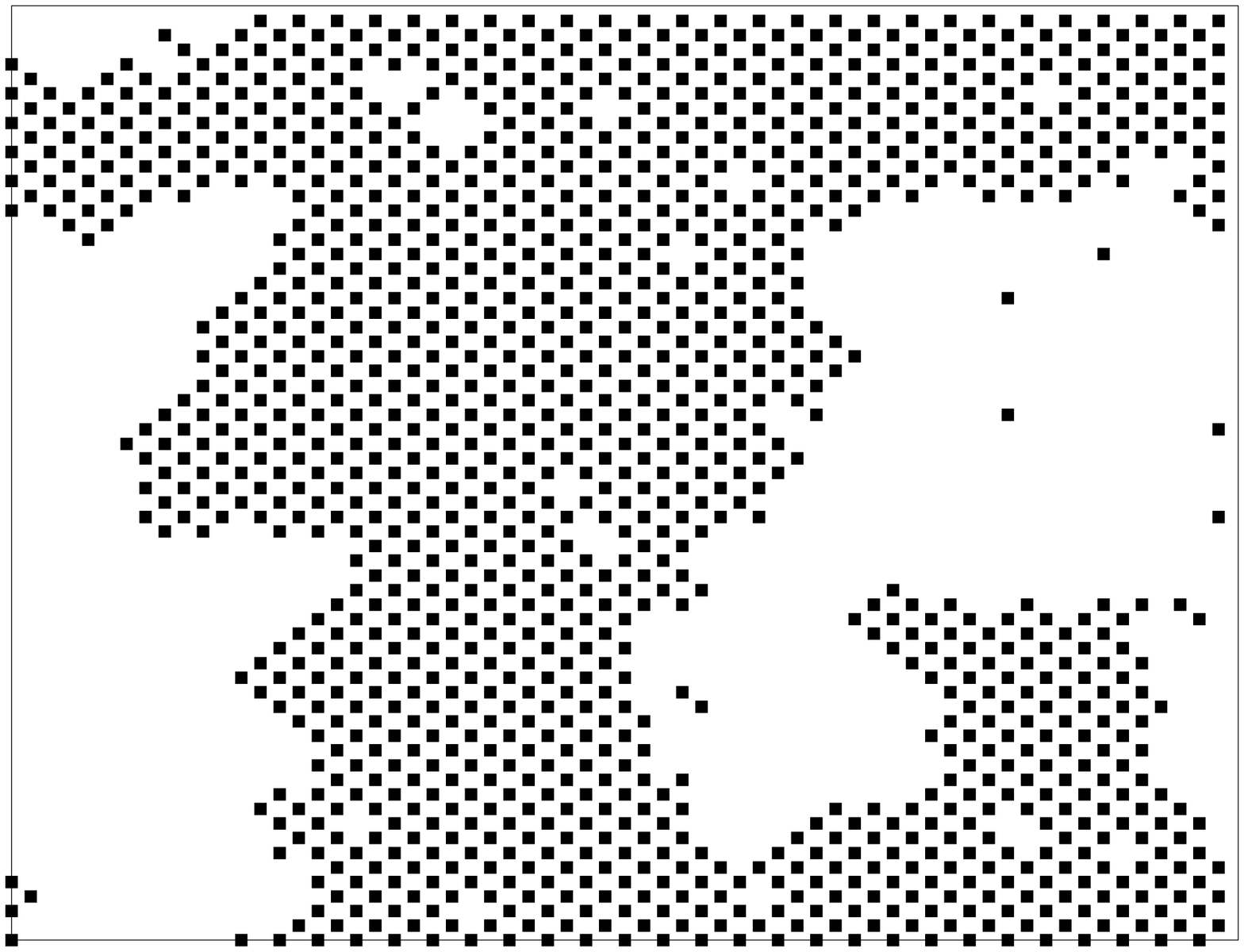}}\quad
\subfigure[]{\includegraphics[bb= 0 0 600 800, scale=0.3, angle= 270]{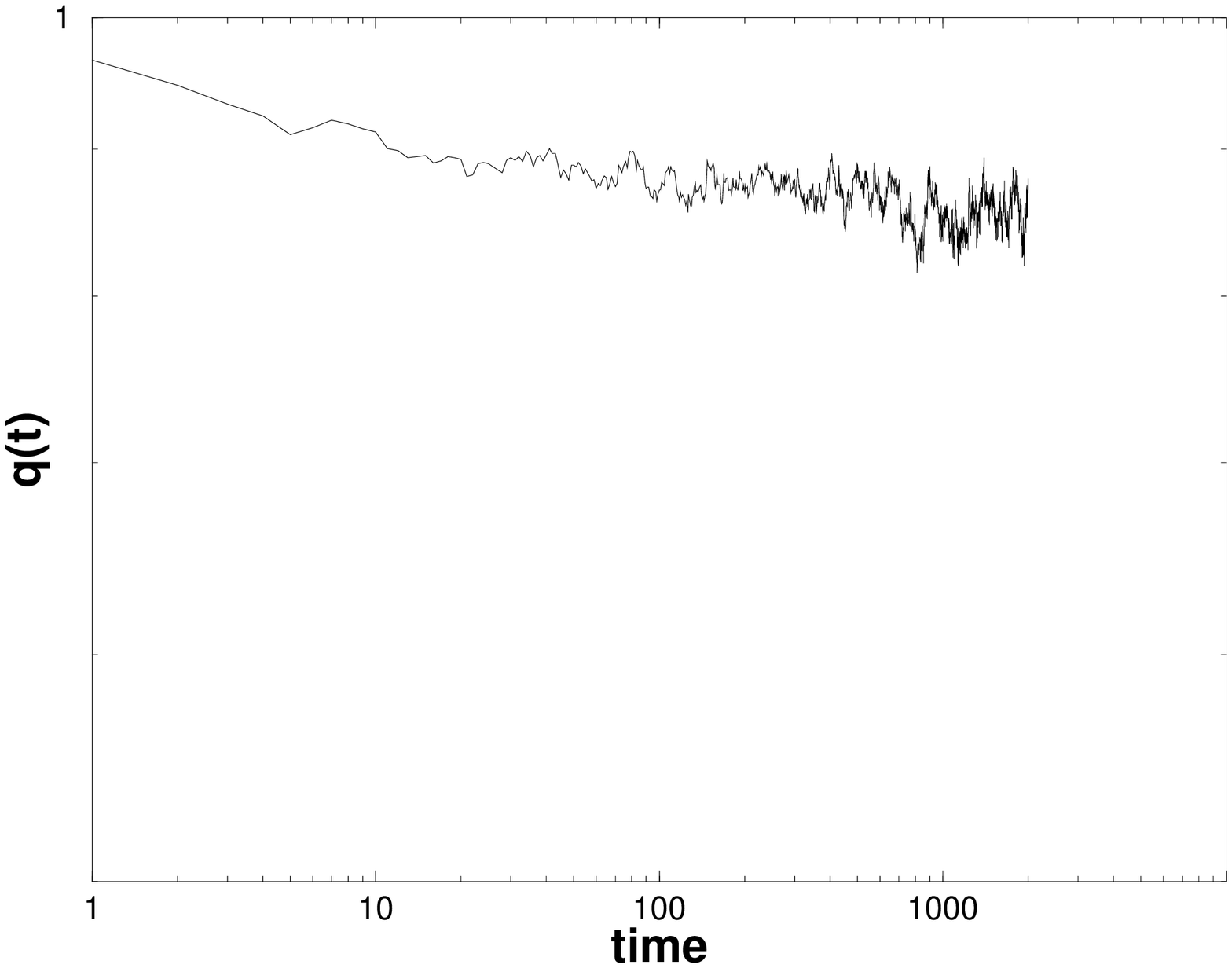}}
\end{center}
\caption{ a) Snapshot of a sample where the spins on one of the sub-lattices
were frozen and the spins on the other sub-lattice evolved accordingly to
\ref{def.eq}. b) The time dependence of the time correlation function
$q(t)$ for the spins on the $B$ sub-lattice. We can see that the curve
relaxes to a non-zero value in accordance with the RFIM expectations.}
\label{fixed.fig}
\end{figure}
}

This result shows that the $B$ domains do not behave as spins in a
traditional spin-glass model, each spin points to a given preferential
direction for all temperatures, in total disagreement with the traditional
spin-glass model where the spin orientation is random at all temperatures
except at $T=0$.

Looking at the snapshot presented on figure \ref{fixed.fig} we can clearly
see that the domain size is much bigger than $\sim 7$, the value proposed
on reference \cite{parisi1998}. Sepp\"al\"a {\it et al} have made exact
zero temperature configuration calculations for the RFIM
\cite{seppala1998}. They define a ferromagnetic break up length scale
$L_b$. For $L=L_b$ the RFIM ground state has a probability of $0.5$ to be
purely ferromagnetic; for larger $L$ this probability decreases and the
ground state magnetization tends to zero. However up to a critical value of
the random field, $\Delta_c$, there will always be a percolating domain
(whose weight tends to zero in the thermodynamic limit). $\Delta$ is
defined as the root mean square random field in units of the ferromagnetic
interaction $J$.  For our model with the $A$ sub-lattice frozen, $\Delta = 2
\lambda$. From  the data presented in \cite{seppala1998}, we can estimate
that for $\lambda = 0.5, 0.7$ and 1.5 (the three cases we will discuss
below), $L_b \sim 45, 22$ and $5$. For total lattice size $L$ the
sub-lattice size is $L/\sqrt{2}$, so that for samples of size $L=64$ the
sub-lattices are close to,  above, and well above $L_b$ respectively for
these three $\lambda$ values.  $\Delta_c$ corresponds to $\lambda = 1$.
For the ground states of the present model with both sub-lattices free, the
value of $L_b$ may not be quite the same as for the pure RFIM. However for
the $\lambda=0.5$ data we will discuss below, we can expect many of the
exact ground states to have complete sub-lattice ferromagnetic ordering at
zero temperature, except for the largest sizes. At finite temperatures
however the spin configurations in thermodynamic equilibrium will have
domains that are large but smaller than the ground state domains.

Now turn to the full model with non-zero $\lambda$ where both sub-lattices
are free. For high enough temperatures the ferromagnetic domains on each
sub-lattice are unstable. At low temperatures the $A$ and $B$ sub-lattices
will conspire so that each induces random fields on the other such that the
total energy is minimized. We will show evidence below that for $\lambda$
up to about 1 there is a low temperature state with frozen large sub-lattice
ferromagnetic domains. For higher $\lambda$ the system appears to be
paramagnetic at all temperatures, like the standard 2d ISG.

\section*{Criteria for an ordering temperature}

There are a number of different criteria which have been used in numerical
work to determine the value of ordering temperatures in spin glasses and
other complex Ising systems.

Finite size scaling on the ``spin glass susceptibility'' is one of these. The
spin glass susceptibility is defined by \begin{equation}
\chi=L^d \langle q^2 \rangle
\end{equation}
where $ \langle q^2 \rangle $ represents the second moment of the
equilibrium autocorrelation function fluctuations and $\langle \ldots
\rangle$ represents both a configurational and thermal average. If
corrections to finite size scaling are negligible, the spin glass
susceptibility follows a scaling rule \cite{bhatt1988} \begin{equation}
\chi(L,T)=L^{d-2+\eta}\, \mbox{f}\,(L^{1/\nu} (T-T_g)), \end{equation}
meaning that precisely at $T_g$, $\log (\chi(L,T_g))$ plotted against
$\log(L)$ should give a straight line of slope $d-2+\eta$. At higher
temperatures $\chi (L)$ should saturate with increasing $L$ while at
temperatures below $T_g$ the log-log plot should curve upwards.
In fact this autocorrelation function susceptibility will show critical
behavior in general at a critical temperature, including cases like
ferromagnets which have standard order parameters. It is thus a parameter
which can be used to identify a transition without the need to specify the
exact nature of the transition.

A complementary finite size scaling method was introduced by Binder
\cite{binder1981}. The dimensionless Binder cumulant \begin{equation}
g_L =\frac{1}{2}\left[ 3- \frac{\langle q^4\rangle}{ \langle q^2\rangle^2}
\right]
\end{equation}
is a parameter characteristic of the shape of the distribution $P(q)$ of
the equilibrium autocorrelation function fluctuations for a given sample of
size $L$ and temperature $T$. For a system with a single characteristic
length scale, $g_L$ is size independent at the ordering temperature, and a
scaling rule applies: \begin{equation}
g_L(T-T_c)=g(L^{1/\nu} (T-T_c)).
\end{equation}
Plots of $g_L(T)$ for different sizes $L$ should all intersect at $T_g$.
This criterion has been widely used, particularly in the spin glass
context. In this model it is also useful to consider the magnetization
Binder parameter defined in the same way: \begin{equation}
g_{mL} =\frac{1}{2}\left[ 3- \frac{\langle m^4\rangle}{ \langle
m^2\rangle^2} \right].
\end{equation}

The time dependence of the autocorrelation function $q(t)$ provides a
further and fundamental criterion for an ordering temperature. In the
thermodynamic limit, for the paramagnetic state in zero external field,
$q(t)$ will always tend to zero at long $t$. As pointed out by Edwards and
Anderson \cite{edwards1975}, if $q(t)$ tends to a finite long time limit
the system can be considered to be ordered; the limiting value of $q(t)$ at
temperature $T$ is the Edwards-Anderson ordering parameter. For a
continuous phase transition, the time scale characteristic of the decay of
$q(t)$ will diverge as the critical temperature is approached from above.
This criterion was used to estimate $T_g$ accurately in the 3 dimension
Ising spin glass measurements of Ogielski \cite{ogielski1985}, who assumed
a standard finite ordering temperature scaling for the characteristic
relaxation time $\tau(T)$. The $T_g$ value defined in this way has been
confirmed later by independent finite size scaling methods.

Marinari et al \cite{marinari1994} found that for the 3d Ising spin glass
the temperature dependence of the spin glass susceptibility could be fitted
equally well by  3 parameter expressions corresponding either to finite
temperature or to zero temperature ordering. It appears that the
temperature dependence of the relaxation time is much more discriminating
than that of the temperature dependence of the susceptibility.

\section*{Simulation techniques and Data}

In order to answer the question of whether the freezing temperature is
finite or not, we have made further simulations for $\lambda=0.5, 0.7$ and
1.5. Wherever direct comparisons could be made, our data are in full
agreement with those of Parisi et al \cite{parisi1998}. Simulations were
carried out using sequential heat bath updating. Samples up to size $64^2$
were studied. Table \ref{samples.tab} present the maximum annealing time
and the number of different realizations for each size. The criterion used
to determine if thermal equilibrium had been attained was the saturation of
the two replica overlap as a function of anneal time \cite{bhatt1988}.

\ifthenelse{\equal{\version}{condmat}}{
\begin{table}
\begin{tabular}{| c | c | c | } 
L & Samples & Anneal Time \\ \hline
4  & 10000 &  15000     \\
8  & 10000 &  15000     \\
16 & 1000 &  150000     \\
32 & 500  &  150000     \\
64 & 500  &  150000     \\
\end{tabular}
\caption{The anneal times and the number of different realizations for each
size studied.}
\label{samples.tab}
\end{table}
}

\subsection{Susceptibility}
For convenience we have followed Parisi et al who used a non-standard spin
glass susceptibility defined by
\begin{equation}
\chi_P =L^2\left[\langle q^2 \rangle - \langle |q^2 | \rangle \right]
\end{equation}
(c.f. the standard spin glass susceptibility defined above) In figure
\ref{xi.fig} we present the dependence of the susceptibility $\chi_P$ with
the size $L$ for different temperatures and for $\lambda=0.5$. The figure shows that the data
follow precisely the behavior to be expected for a system with an ordering
temperature lying somewhere between $2.1$ and $2.2$. For higher
temperatures the susceptibility saturates with increasing size; for lower
temperatures the $\log(\chi_P)$ against $\log(L)$ plot curves upwards. At
around $T=2.2$, $\log(\chi_P)$ increases linearly with $\log(L)$, which is
the signature of
critical behaviour. From the slope of the critical line, the exponent
$\eta$ can be estimated to be 1.2$\pm$0.1 .

\ifthenelse{\equal{\version}{condmat}}{
\begin{figure}
\begin{center}
\includegraphics[bb= 0 0 600 800, scale=0.3, angle= 270]{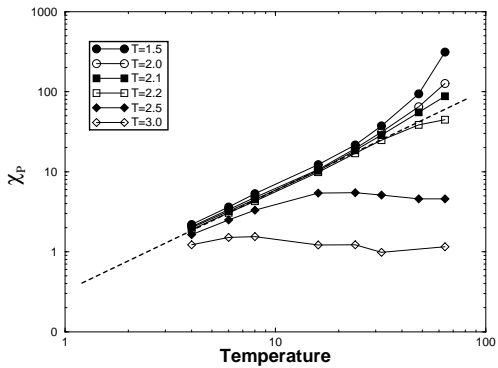}
\end{center}
\caption{ The dependence of the spin glass susceptibility with $L$ for
different temperatures. For $T\sim 2.2$ we have $\chi\sim L^\alpha$. }
\label{xi.fig}
\end{figure}
}

\subsection{Binder Cumulant}

In \cite{lemke1996b} the Binder cumulant crossing method for sample sizes
up to $L=12$ was used to estimate ordering. Parisi et al \cite{parisi1998}
showed that for larger sample sizes the crossing point of the Binder curves
$g_L(T)$ moved to lower temperatures and became badly defined. On the basis
of this observation they suggested that in fact for large sizes there is no
ordering temperature, and that the low temperature state is paramagnetic.

The Binder cumulant method can be delicate to use. This can be illustrated
by a trivial ``paradox'' in the present system. For any standard Ising
ferromagnet, $g_L(T)$ goes to $1$ at low temperatures. However in the
present system, if $\lambda$ is zero (so each of the two sub-lattices order
ferromagnetically), because of the four possible ground states $g_L(T)$
goes to $0.5$ at low temperatures, not to $1$. In the general case,
$g_L(T)$ curves going to small values or even to zero at low temperatures
for large systems is not the signature of a paramagnetic state, but rather
of the system having a large number of orthogonal ground states. The
classical Binder cumulant behavior with a well defined crossing point and
good scaling above and below the critical temperature will be observed for
systems with a single effective correlation length and a standard evolution
with size and temperature for the form of the distribution $P(q)$. Certain
non-standard systems with {\it bona fide} ordering transitions show very
unorthodox behavior for the Binder cumulants \cite{bernardi1999}.

Instead of concentrating attention on the details of the Binder cumulant
curves at the lowest temperatures, we can do trial scaling plots over the
whole temperature range for $\lambda=0.5$. For the scaling plots we can assume:
\begin{itemize}
\item that there is a critical temperature at about $T=2.1$ as
indicated by the susceptibility scaling, or \item that there is a zero
temperature critical point and an exponent
$\nu$ equal to the value obtained from scaling of the Binder cumulant for
data on the standard 2d ISG, i.e. $\nu=1.4\pm 0.2$ \cite{bhatt1988}.
\end{itemize}

The two scaling plots are shown in figure \ref{scale.fig}, while the raw
data is presented on figure \ref{gl.fig}. It can be seen immediately that
the first assumption leads to acceptable scaling, at least above the
assumed critical point. The zero temperature scaling is clearly incorrect.
We conclude that the finite critical temperature scaling is compatible with
the Binder cumulant data while the zero temperature scaling is not.

\ifthenelse{\equal{\version}{condmat}}{
\begin{figure}
\begin{center}
\subfigure[]{\includegraphics[bb= 0 0 600 800, scale=0.3, angle= 270]{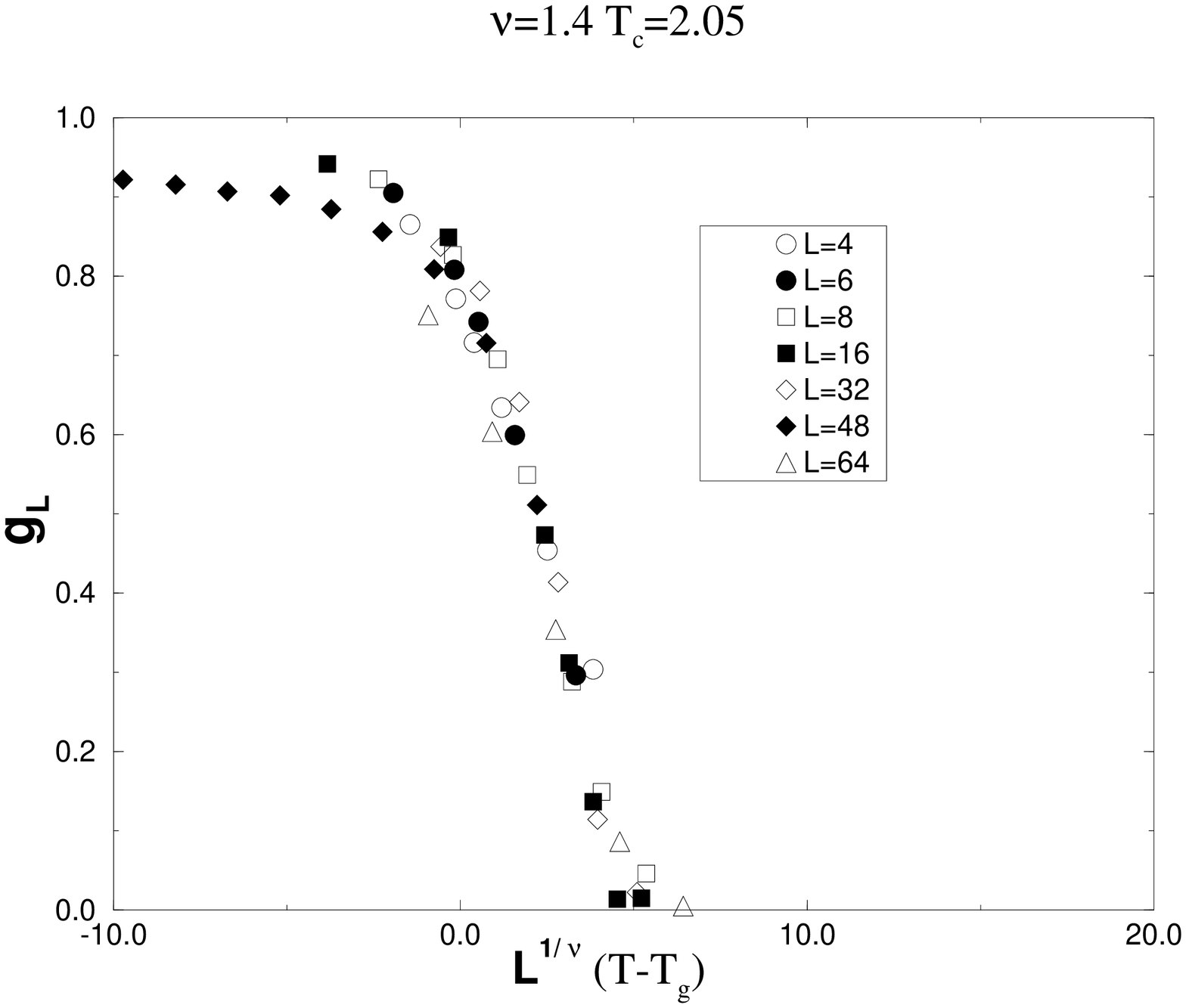}}\quad
\subfigure[]{\includegraphics[bb= 0 0 600 800, scale=0.3, angle= 270]{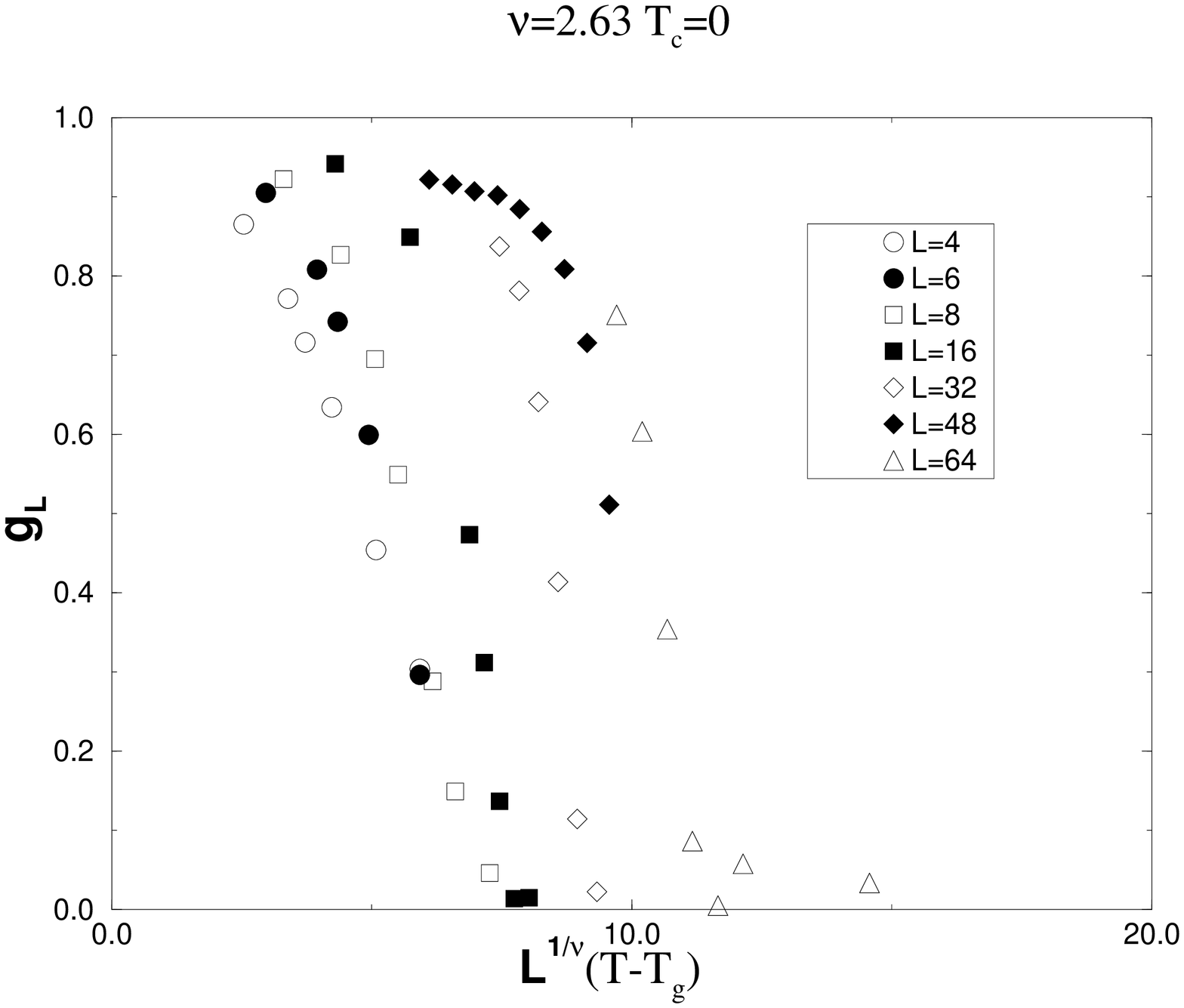}}
\end{center}
\caption{ In figure a) the scaling proposed by reference Parisi et al in
b) the scaling obtained by supposing $T_g=2.05$ and $\nu=1.4$. The data 
for $L=48$ were obtained from Parisi {\it et al}.} 
\label{scale.fig}
\end{figure}

\begin{figure}
\begin{center}
\includegraphics[bb= 0 0 600 800, scale=0.3, angle= 270]{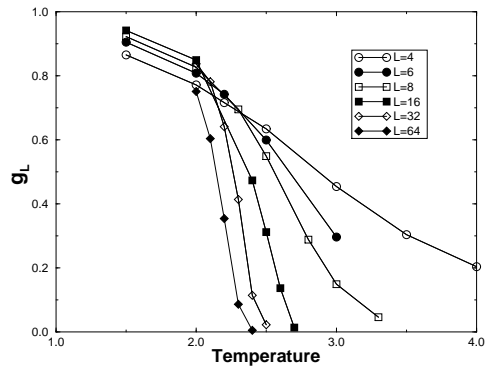}
\end{center}
\caption{ The dependence of the Binder parameter with $L$ for different
temperatures.}
\label{gl.fig}
\end{figure}
}

The fact, underlined by Parisi et al, that the low temperature Binder
cumulant values do not increase regularly with increasing $L$ does not
indicate that the low temperature state is paramagnetic, but that the type
of order at low temperatures is evolving as $L$ increases. Direct
evaluation shows that for $L=4$ there are frequently just two ground
states, a purely ferromagnetic or antiferromagnetic state and its mirror
image. This ``staggered ferromagnetism'' is what would be expected from the
discussion of the RFIM given above; the sub-lattice magnetizations are
essentially ferromagnetic and in any particular sample the random
interactions select a ferro or an anti-ferro coupling between sub-lattices.
In consequence for these small samples $g_L$ must tend to exactly $1$ at
zero temperature and will be close to 1 at higher temperatures. This
behavior will continue to hold until sizes are reached where $L$ is of the
order of the $L_b$ at the particular temperature studied. If for larger
samples there are many alternative more complex Gibbs states at that
temperature, $g_L(T)$ will become lower for larger $L$. This seems to be
the real situation, with temperature dependent crossover sizes.

The sub-lattice magnetism Binder cumulant tends to $\sim 1$ at temperature$T
\sim 1.5$ for samples up to size $L=8$, and then decreases regularly with
increasing sample size \cite{parisi1998} figure 4. This indicates that the
sub-lattices are ferromagnetic in the small samples, and in the larger
samples each sub-lattice is principally either up or down (not zero
magnetization as for large samples in the strict RFIM) but contain domains
of non majority spin, at least at finite temperatures. As $L$ increases the
average sub-lattice magnetization drops, but it would need very large $L$
for the sub-lattice magnetization distribution to take up a Gaussian form
centered on zero \cite{seppala1998}. This gradual evolution with sample
size, most clearly observed in the sub-lattice magnetization cumulant, is
certainly also the cause of the ``anomalous'' low temperature behaviour of
the sublattice $g_L$ cumulant and the global $g_L$ cumulant at low
temperatures (see figures 7 and 9 of reference \cite{parisi1998}). From the
discussion above, in these particular systems we can expect deviations from
asymptotic large scale behaviour until very large values of $L$, well
beyond the values used so far in the simulations.

Finally, it must be remembered that a Binder cumulant is not directly
sensitive to whether the spins are frozen or not.

\subsection{Relaxation}

We measured the autocorrelation function decay $q(t)$ after long
equilibration anneals at different temperatures for large samples, $L=64$.
This was done for $\lambda =0.5, 0.7$, and $1.5$; the results are presented
on figures \ref{qA.fig}, \ref{qB.fig} and \ref{qC.fig}. At each temperature
the form of the decay can be seen to be initially algebraic $q(t) \sim
t^{-x}$, with a cutoff function at longer times. For the first two values
of $\lambda$, as $T$ is reduced towards a temperature close to 2.1, the
relaxation becomes purely algebraic to long time scales, meaning that the
characteristic time defining the cutoff function is diverging. The
characteristic time for the decay can be defined either by $\tau_c$ or by
$\tau_{av}$ where \begin{eqnarray}
\tau_c &=& \int_0^\infty q(t), \\
\tau_{av} &=& \frac{\int_0^\infty tq(t)} {\int_0^\infty q(t)} \\ \end{eqnarray}
$\tau_c$ and $\tau_{av}$ were calculated for convenience by fitting the
$q(t)$ curves with an Ogielski function \cite{ogielski1985} :
\begin{equation}
q(t)=t^{-x}\exp\left[ -\left(\frac{t}{\tau}
\right)^\beta\right] .
\end{equation}

\ifthenelse{\equal{\version}{condmat}}{
\begin{figure}
\begin{center}
\includegraphics[bb= 0 0 600 800, scale=0.3, angle= 270]{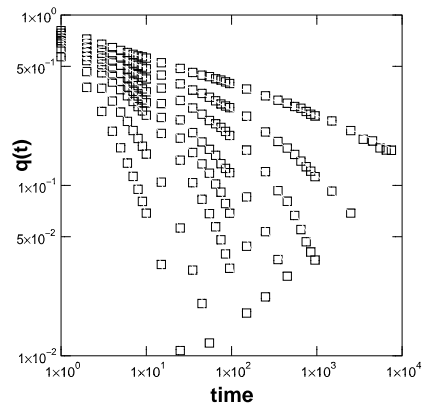}
\end{center}
\caption{ The time dependence of the correlation function for $\lambda=0.5$ and different
temperatures. From bottom to top $T=$ 3.5, 3.0, 2.7, 2.6, 2.5, 2.4, 2.3, 2.2.}
\label{qA.fig}
\end{figure}

\begin{figure}
\begin{center}
\includegraphics[bb= 0 0 600 800, scale=0.3, angle= 270]{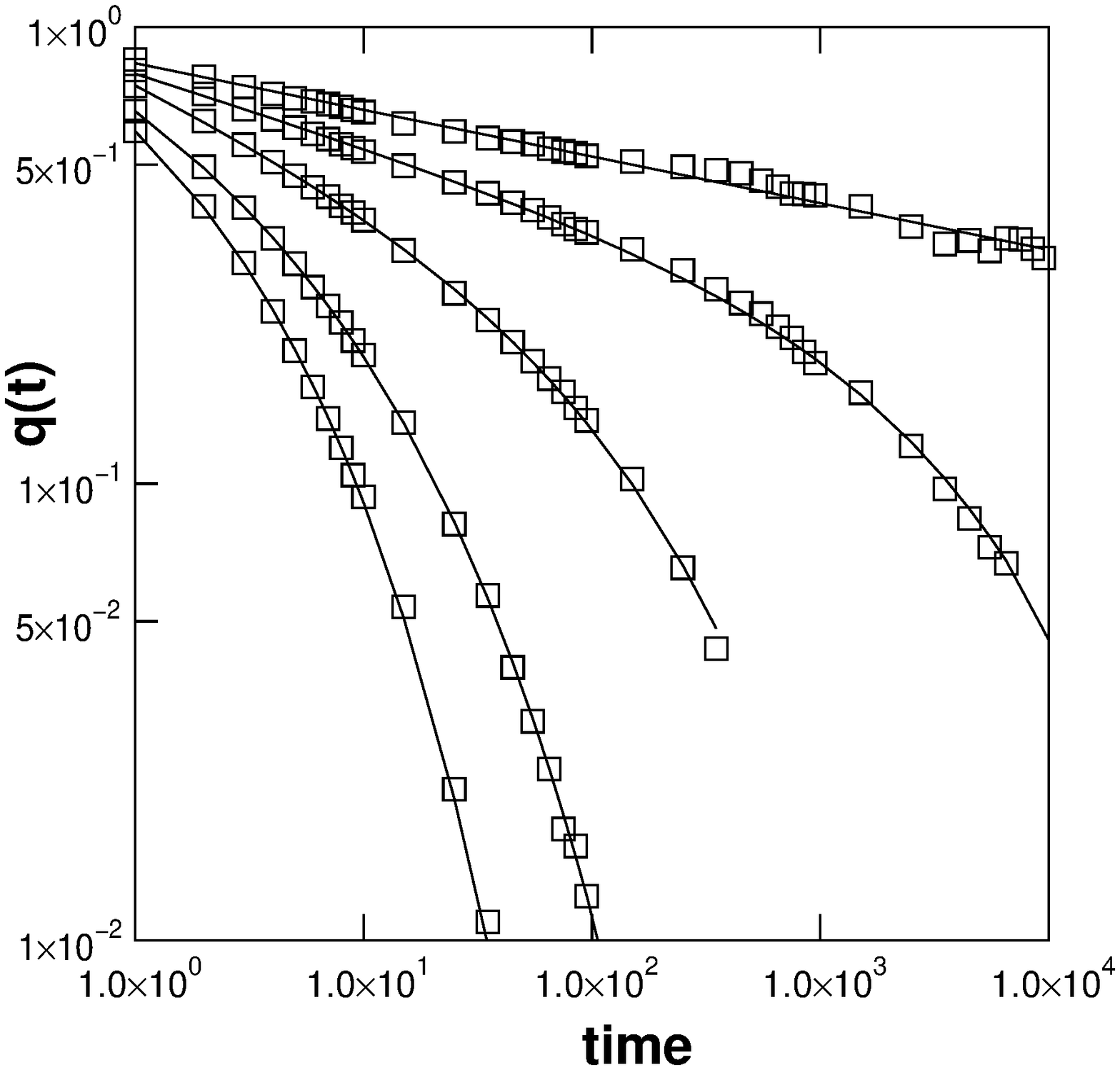}
\end{center}
\caption{ The time dependence of the correlation function for $\lambda=0.7$ and different
temperatures. From bottom to top $T=$ 3.5, 3.0, 2.5, 2.3, 2.2, 2.0.}
\label{qB.fig}
\end{figure}

\begin{figure}
\begin{center}
\includegraphics[bb= 0 0 600 800, scale=0.3, angle= 270]{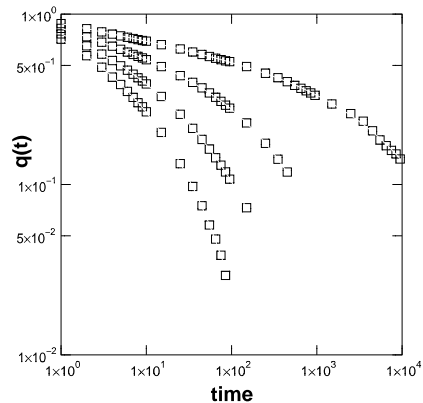}
\end{center}
\caption{ The time dependence of the correlation function for $\lambda=1.5$ and different
temperatures. From bottom to top $T=$ 3.5, 3.0, 2.5, 2.0.} \label{qC.fig}
\end{figure}
}

For $\lambda =0.5$, $\tau_c(T)$ and $\tau_{av}(T)$ diverge at a temperature
just below $T=2.1$, figure \ref{tau.fig}. The critical value of the
exponent $x$ at the temperature where $\tau(T)$ diverges is about 0.15. For
$\lambda =0.7$ the behaviour is very similar and the exponent is about
0.11. This means that for both these values of $\lambda$, below a critical
temperature and in the thermodynamic limit the system is frozen to
indefinitely long times with a finite Edwards Anderson parameter, and so it
is ordered (in the Edwards Anderson sense). We can note that the behaviour
is very similar for these two values of $\lambda$, although from the
discussion above we can expect the latter to have a break up length $L_b$ half as
large as in the former.For $\lambda = 0.5$ the zero temperature $L_b$ estimated above is
about equal to the sublattice size (which we can take equal to $L/\sqrt(2)$)
while for $\lambda =0.7$ $L_b$ is significantly smaller than the lattice
size. This criterion does not appear to play a major role at the temperature
where the relaxation time is diverging.

\ifthenelse{\equal{\version}{condmat}}{
\begin{figure}
\begin{center}
\includegraphics[bb= 0 0 600 800, scale=0.3, angle= 270]{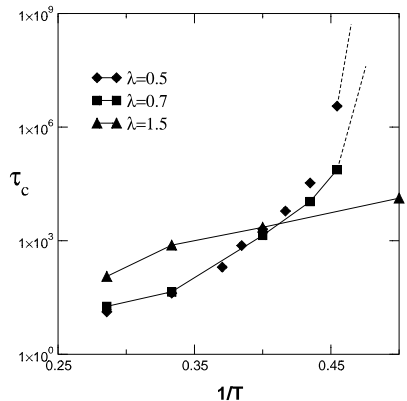}
\end{center}
\caption{ The temperature dependence of $\tau_c$ with
temperature for $\lambda$ =0.5, 0.7, 1.5. The figure shows clearly that
$\tau_c$ is diverging for $\lambda$ =0.5 and 0.7, and grows slowly for
$\lambda$ =1.5. }
\label{tau.fig}
\end{figure}
}

The behaviour for $\lambda=1.5$ is quite different; the temperature
variation of the relaxation time is very much slower, figure \ref{qC.fig}
and \ref{tau.fig}. In this case the form of $\tau (T)$ can be compared with
that seen in the standard 2d ISG case, where $\tau(T)$ is seen to diverge
only as zero temperature is approached, following an Arrhenius like law
\cite{mcmillan1983}. The $q(t)$ measurements are clearly discriminatory and
can distinguish systems with finite freezing temperatures from ones with
zero temperature (or at least very low temperature) ordering.

The susceptibility scaling, Binder cumulant scaling, and autocorrelation
relaxation data thus give conclusive and consistent evidence for freezing
at a temperature near 2.1 for $\lambda=0.5$. The relaxation data indicate a
slightly lower freezing temperature for $\lambda =0.7$, and a much lower
temperature freezing compatible with $T_g=0$ for $\lambda= 1.5$. The
estimated freezing temperatures are very similar to those suggested
originally in \cite{lemke1996b}.  There seems to be no evidence for an
onset of paramagnetic behaviour at low temperatures with increasing size.

We conclude that the low temperature state is frozen for $\lambda$ values
up to about 1. It is perhaps not a coincidence that the critical value of
$\lambda$ appears to correspond to the critical RFIM $\Delta_c$ as defined
above.

Can the present system be described as a spin glass ? We can attempt to
give a coherent description of the low temperature frozen state in the
light of the different types of data. As we have seen, for small sizes the
ordering can indeed be described in terms of ``staggered ferromagnetism''.
For the larger sizes covered in this work and in \cite{parisi1998}, the low
temperature sublattice magnetism Binder cumulant decreases regularly with
increasing $L$, but is still as high as 0.8 at $L=48$ and $T=1.5$.
(\cite{parisi1998} figure 4). This indicates that in equilibrium at this
temperature, each sublattice is split up into fairly big ferromagnetic
domains with magnetization of both signs, but for each particular replica,
one sign of magnetization is preponderant for each sublattice. However
within each sublattice there are large minority domains. We have found that
if a large sample is cooled a number of different times to a temperature
below the critical temperature estimated above, the quasi-stationary
pattern of domains observed is far from identical each time (in contrast to
what is always seen in the standard RFIM). We can extrapolate, and surmise
that for very large $L$ there would be no magnetization bias for a
sublattice, and there would then be for the entire system a very large
number of possible Gibbs states below the ordering temperature, nearly
orthogonal to each other in phase space. The whole system can be understood
as freezing at low temperatures because once domains of a maximal size are
formed, the domain walls are pinned by the effect of the random
interactions. The low temperature state would then ressemble a spin glass
in that there are many Gibbs states, but the local spin structure is
entirely different because of the strong local ferromagnetic correlations
within each sublattice.

\section*{Conclusions}

We have investigated in more detail the ferromagnetic plus random
interaction system described by equation (\ref{def.eq}). To summarize:
data on autocorrelation function or ``spin glass'' susceptibility, Binder
cumulant, and autocorrelation function relaxation, all consistently
indicate a critical temperature for freezing of $T_c \sim 2.1$ for $\lambda
=0.5$. Relaxation data indicate a slightly lower freezing temperature for
$\lambda =0.7$, and are compatible with a zero temperature freezing for
$\lambda =1.5$.  Therefore in the range of $\lambda$ up to about 1, this
two dimensional RCF system with interactions which are partly random has
a finite freezing temperature. There is no evidence for a return to
paramagnetic behaviour (with faster relaxation for instance) with
increasing size. Independent defect free energy data confirm our
conclusions \cite{shirakura1998}. The finite  freezing temperature result is not
in contradiction with the general consensus that for standard 2D spin glass
model the ordering temperatures are either zero or at least very low. The
picture for the transition suggested above implies that the transition
mechanism for this system is radically different from that of the standard
spin glass. We have no reason to expect that this system can be mapped onto
a standard Ising spin-glass, even though this system shares many properties
with the traditional model: frustration, complex phase space landscape etc.
Because of the ferromagnetic short range ordering within each sublattice,
the term ``cluster glass'' would probably be more appropriate than ``spin
glass''.

Many interesting questions remain; in particular it will be very important
to describe accurately the nature of the low temperature phase, and to
obtain explicity information about the domain size distribution, and the
domain geometry characteristics in large samples and at low temperatures.
It should be possible to apply sophisticated  methods to establish ground
state characteristics.

Finally we believe
this model should be a very useful laboratory to test theoretical issues
concerning disordered systems, since in this case we have a freezing
transition in a two-dimensional system, where theoretical analysis, exact
ground state methods, simulations, and visualization techniques are easier
to apply than in higher dimensions. Further progress would however require
the study of much larger samples.

\acknowledgments
We would like to thank T. Shirakura for showing us his unpublished
data. N. L. would like to thank the kind hospitality of the Laboratoire de Physique des Solides during the
preparation of this manuscript.

\bibliographystyle{prsty} 

\ifthenelse{\equal{\version}{referee}}{

\newpage

\section*{Tables}

\begin{table}
\begin{tabular}{| c | c | c | } 
L & Samples & Anneal Time \\ \hline
4  & 10000 &  15000     \\
8  & 10000 &  15000     \\
16 & 1000 &  150000     \\
32 & 500  &  150000     \\
64 & 500  &  150000     \\
\end{tabular}
\caption{The anneal times and the number of different realizations for each
size studied.}
\label{samples.tab}
\end{table}

\newpage

\section*{Figures}
\begin{figure}
\caption{ a) Snapshot of a sample where the spins on one of the sub-lattices
were frozen and the spins on the other sub-lattice evolved accordingly to
\ref{def.eq}. b) The time dependence of the time correlation function
$q(t)$ for the spins on the $B$ sub-lattice. We can see that the curve
relaxes to a non-zero value in accordance with the RFIM expectations.}
\label{fixed.fig}
\end{figure}

\begin{figure}
\caption{ The dependence of the spin glass susceptibility with $L$ for
different temperatures. For $T\sim 2.2$ we have $\chi\sim L^\alpha$. }
\label{xi.fig}
\end{figure}

\begin{figure}
\caption{ In figure a) the scaling proposed by reference Parisi et al 
b) the scaling obtained by supposing $T_g=2.05$ and $\nu=1.4$. The data 
for $L=48$ were obtained from Parisi {\it et al}.} 
\label{scale.fig}
\end{figure}

\begin{figure}
\caption{ The dependence of the binder parameter with $L$ for different
temperatures.}
\label{gl.fig}
\end{figure}

\begin{figure}
\caption{ The time dependence of the correlation function for different
temperatures. From bottom to top $T=$ 3.5, 3.0, 2.7, 2.6, 2.5, 2.4, 2.3, 2.2.}
\label{qA.fig}
\end{figure}

\begin{figure}
\caption{ The time dependence of the correlation function for different
temperatures. From bottom to top $T=$ 3.5, 3.0, 2.5, 2.3, 2.2, 2.0.}
\label{qB.fig}
\end{figure}

\begin{figure}
\caption{ The time dependence of the correlation function for different
temperatures. From bottom to top $T=$ 3.5, 3.0, 2.5, 2.0.} \label{qC.fig}
\end{figure}

\begin{figure}
\caption{ The temperature dependence of $\tau_c$ with
temperature for $\lambda$ =0.5, 0.7, 1.5. The figure shows clearly that
$\tau_c$ is diverging for $\lambda$ =0.5 and 0.7, and grows slowly for
$\lambda$ =1.5. }
\label{tau.fig}
\end{figure}
}

\end{document}